\documentclass[10pt,twocolumn,prl]{revtex4-1}
\usepackage{graphicx}
\usepackage{amssymb}
\usepackage{amsmath}

\begin{document}

\title{
Metal nanoparticles with sharp corners: Universal properties of plasmon resonances }

\author{B.\ Sturman$^1$, E.\ Podivilov$^1$, and M.\ Gorkunov$^2$}
\affiliation{\vspace*{1mm}$^1$Institute of Automation and Electrometry, Russian Academy
of Sciences, 630090 Novosibirsk, Russia \\
$^2$Shubnikov Institute of Crystallography, Russian Academy of Sciences, 119333 Moscow,
Russia}


\begin{abstract}

We predict the simultaneous occurrence of two fundamental phenomena for metal nanoparticles possessing sharp corners: First, the main plasmonic dipolar mode experiences strong red shift with decreasing corner curvature radius; its resonant frequency is controlled by the apex angle of the corner and the normalized (to the particle size) corner curvature. Second, the split-off plasmonic mode experiences strong localization at the corners. Altogether, this paves the way for tailoring of metal nano-structures providing wavelength-selective excitation of localized plasmons and a strong near-field enhancement of linear and nonlinear optical phenomena.

\vspace*{1mm}
PACS numbers: 73.20.Mf, 42.25.Fx, 78.40.Kc

\end{abstract}

\maketitle

Plasmon excitations of metal nanoparticles, including 2D particles (nanowires), is a vast and hot research area. Potential applications of nano-plasmons span from nano-lasers (spasers) \cite{S1,NL1,NL2} to bio-sensors and sensors of single atoms and molecules \cite{Schuller10,Gissen11}. It is highly important to apprehend the possibilities of tailoring of plasmonic resonances and the extends of concentration of the light energy deeply on the sub-wavelength scale.

Analytical plasmonic solutions to Maxwell's equations are available only for metal nanoparticles of simplest 2D and 3D shapes -- like circular cylinder, sphere, and ellipsoid~\cite{Bohren,Novotny07}. They include the optical permittivity of the metal $\varepsilon_{\rm M}(\omega) = \varepsilon'_{\rm M}(\omega) + i\varepsilon''_{\rm M}(\omega)$, such that $\varepsilon'_{\rm M} < 0$ and $\varepsilon''_{\rm M} \ll |\varepsilon'_{\rm M}|$, and give often strongly degenerated plasmonic resonances. In particular, all plasmonic excitations of a circular cylinder correspond to a single eigenfrequency given by $\varepsilon'_{\rm M}(\omega) =~-1$.

Using direct numerical methods, serious efforts have been undertaken to ascertain the phenomenology of the plasmonic response of 2D and 3D particles of more complicated shapes~\cite{Ruppin,Kottmann2,Kelly03,Gonzalez07}. The general outcome is that the deviations from the most symmetric shapes result in enriching plasmon spectrum and in the appearance of new red-shifted resonances. In the 2D case, the strongest shape effects were found for triangular cross-sections.

Regardless of the plasmonic effects, the presence of sharp tips and corners (metal or dielectric) is known to lead to the corner singularities, i.e., to a strong local enhancement of electromagnetic fields~\cite{ECM,Corner1}. At the same time, the impact of tips and corners on the plasmonic properties of nanoparticles remains in essence unexplored. In numerical simulations, the sharp surface features are typically rounded to avoid numerical instabilities. Non-rounded $\pi/2$-corners lead to divergence of numerical methods for $\varepsilon_{\rm M} \to -3$~\cite{WE}. Mathematical foundations of the plasmonic theory~\cite{Kellogg,Mikhlin} refer to sufficiently smooth (Lyapunov) surfaces. A recent heuristic attempt to deal with plasmonic properties of nanoparticles possessing perfectly sharp features~\cite{Vincent} is strongly unconvincing.

Experimentally, the presence of sharp features of metal nanoparticles is far from exotic. Modern chemical methods for production such particles are able to provide almost atomically sharp edges in the crystallographic directions~\cite{Chem1,Chem2,Chem3}. Synthesis of metal nanostars has allowed recently to strongly enhance the near fields at the plasmonic resonance~\cite{Nanostars1}.

In this Letter, we predict and analyze a strong impact of corners of metal nanoparticles on fundamental plasmonic properties. Both the spectrum and the spatial structure of the plasmonic modes are strongly affected by the corners showing simple and universal features.

We employ the most efficient approach to the description of plasmonic resonances -- the method of integral equations for the amplitude of the surface-charge density $\sigma({\bf r})$~\cite{PhilMag89,Mayergoyz05}. Applicable in the quasi-static approximation, it treats the plasmonic eigenproblem as a geometric issue: The real resonant values of the optical permittivity $\varepsilon_j$ depend only of the shape of the particle, while the corresponding eigenfunctions $\sigma_j({\bf r})$ are scaling invariant. The knowledge of the eigenmodes allows to determine the width of the resonance, the polarizability of the particle, near fields, etc., using a perturbation routine. Reduction of the dimension of the problem to be solved strongly enhances the capability of numerical methods.

In the 2D case, the eigenproblem to be solved reads:
\begin{equation}\label{2DEquation}
\int_L K({\bf r},{\bf r}')\,\sigma_j({\bf r}') \,dl' = \Lambda_j\,
\sigma_j({\bf r})\;, \qquad {\bf r},{\bf r}' \in L \;,
\end{equation}
where the line $L$ is the boundary of the particle, $dl$ is the length element along $L$ and $\Lambda_j = (\varepsilon_j + 1)/(\varepsilon_j - 1)$ is the eigenvalue. The kernel $K$ is real and given by
\begin{equation}\label{2DKernel}
K({\bf r},{\bf r}') = \frac{{\bf n} \cdot ({\bf r} - {\bf r}')}{\pi \,({\bf r}
- {\bf r}')^2} \;,
\end{equation}
where ${\bf n} = {\bf n}({\bf r})$ is the unit vector of the external normal. Generally, Eq.~(\ref{2DEquation}) gives an infinite sequence of the modes.

The integral eigenproblem for $\varepsilon_j$ and $\sigma_j({\bf r})$ possesses important general features~\cite{Mayergoyz05}. The total modal charge is zero, $\int_L \sigma_j({\bf r})\,dl = 0$. Since the kernel is not symmetric, $K({\bf r},{\bf r}') \neq K({\bf r}',{\bf r})$, the eigenproblem is not Hermitian. Thus, one has to employ additionally the eigenfunctions $\tau_j({\bf r})$ of the adjoint problem with the transposed kernel $K({\bf r}',{\bf r})$ and the orthogonality relation $\int_L \sigma_j({\bf r})\,\tau_{j'}({\bf r})\,dl = 0$ for $j \neq j'$. In the 2D case, the eigenvalues appear in twin pairs $\varepsilon_j$, $\varepsilon_j^{-1}$.

\begin{figure}[h]
\centering
\includegraphics[width=8cm]{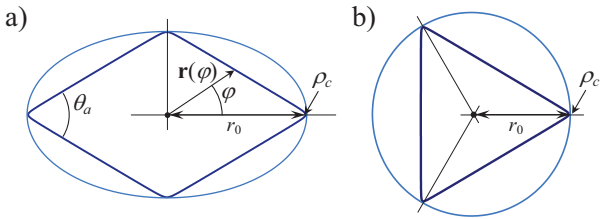}
\caption{The boundary line $L$ for a smoothed rhombus transferring to an ellipse (a) and a smoothed equilateral triangle transferring to a circle (b).}\label{Rhombus}
\end{figure}

We specify the boundary line $L$ by the polar-angle dependence of the radius, $|{\bf r}|(\varphi) = r(\varphi)$. To investigate the impact of corners, we employed different parametrizations of $r(\varphi)$ with variable corner-curvature radius $\rho_c$ and apex angle $\theta_a \leq \pi/2$. The simplest one is given by
\begin{equation}\label{parametrization}
r = \frac{r_0(p + 1)}{\sqrt{\cos^2\hspace*{-0.5mm}\varphi + a^2p^2\sin^2\hspace*{-0.5mm}\varphi} + \sqrt{a^2\hspace*{-0.5mm}\sin^2\hspace*{-0.5mm}\varphi + p^2\cos^2\hspace*{-0.5mm}\varphi}} \,,
\end{equation}
where $r_0 = r(0)$ is the large half-diameter, $a = {\rm ctg}(\theta_a/2)$, and $1 \leq p < \infty$. It gives a smoothed rhombus with the sharp apex angle $\theta_a$ and the normalized corner curvature  $\kappa_c = r_0/\rho_c = {\rm ctg}^2(\theta_a/2)(p - 1 + p^{-1})$ growing linearly with $p$ for $p \gg 1$, see also Fig.~\ref{Rhombus}a. The case $p = 1$ corresponds to an ellipse with the axes ratio of $a$. Another parametrization, illustrated by Fig.~\ref{Rhombus}b, is a smoothed equilateral triangle ($\theta_a = \pi/3$) transferring to the circle for $\rho_c = r_0$. The kernel $K({\bf r},{\bf r}')$ can be expressed by $r(\varphi)$, $r(\varphi')$, and $dr/d\varphi$; it peaks at the corners with $K({\bf r}_c,{\bf r}_c) = 1/2\pi \rho_c$.

Equation~(\ref{2DEquation}) with the described boundary lines $L$ was solved numerically using crowding of the points near the corners. It was made sure that further increasing of the number of points does not influence the results. All general properties of $\varepsilon_j$ and $\sigma_j({\bf r})$ were fulfilled.

\begin{figure}[h]
\centering
\includegraphics[width=7.1cm]{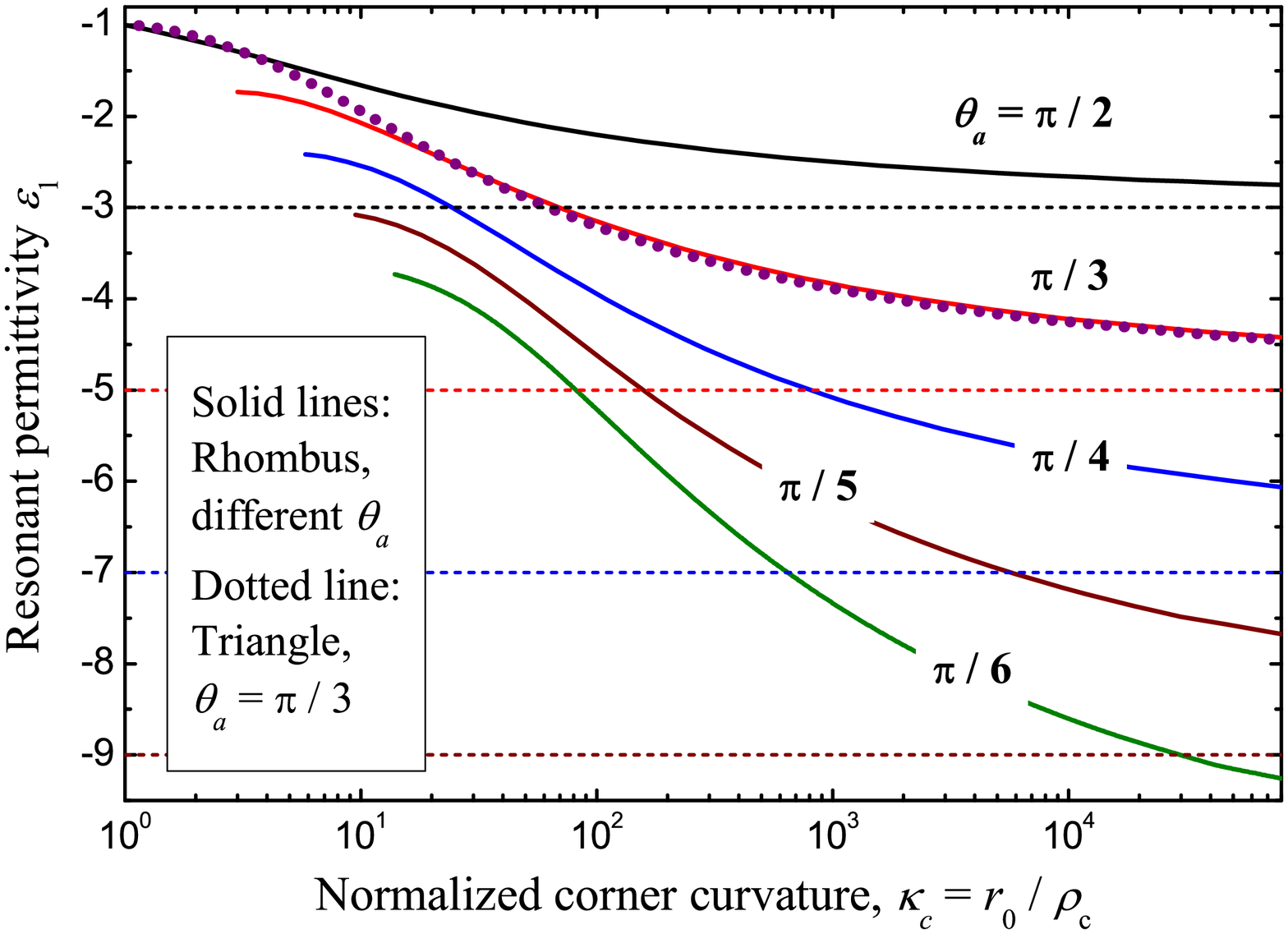}
\caption{The lowest dipolar branch $\varepsilon_1(\kappa_c)$ for the smoothed rhombus with the apex angle $\theta_a = \pi/2,\pi/3,\pi/4,\pi/5$, and $\pi/6$ (the solid lines) and a smoothed equilateral triangle with $\theta_a = \pi/3$ (the dotted line). The horizontal dashed lines show the corresponding limiting values $1 - 2\pi/\theta_a$.}\label{Lower}
\end{figure}

The solid lines in Fig.~\ref{Lower} show the dependences $\varepsilon_1(\kappa_c)$ for the lowest dipolar branch and several values of the apex angle $\theta_a$ of the rhombus on a semi-logarithmic scale. The starting values of $\varepsilon_1$ correspond to the known plasmonic solutions for the elliptic cross-section. Increasing corner curvature $\kappa_c$ results in strong decrease of the resonant permittivities [i.e., in strong red shifts of the resonant frequencies in accordance with the Drude-like dependences $\varepsilon_{\rm M}(\omega)$] for all values of~$\theta_a$. This decrease persists for $\kappa_c \gtrsim 10$, when the shape is already settled down and increasing sharpness of the corners remains the only variable feature. The dotted line is plotted for the triangular parametrization, $\theta_a = \pi/3$. When $\kappa_c$ increases, it approaches quickly the solid line corresponding to rhombus with the same $\theta_a$. For $\kappa_c \to 1$, the dotted line and the solid line for $\theta_a = \pi/2$ tend to $1$: This is the limit of the circular cross-section with $\varepsilon_j = -1$. We have made sure also that different types of smoothing give essentially the same results for $\kappa_c \gg 1$.

The dipolar branches $\varepsilon_j(\kappa_c)$ with $j \geq 2$ are well separated from the main branch $\varepsilon_1(\kappa_c)$ for $\kappa_c \gg 1$, $|\varepsilon_1 - \varepsilon_{2,3}| \approx |\varepsilon_1 + 1|$. This makes possible selective excitation of the main branch. At the same time, the vicinity of the point $\varepsilon = -1$ is always filled up with the spectrum.

The horizontal dashed lines in Fig.~\ref{Lower} indicate the critical values $\varepsilon_c = 1 - 2\pi/\theta_a$ which correspond to non-integrable field singularities for a perfect single corner with the apex angle $\theta_a$~\cite{Corner1}. Obviously, the branch $\varepsilon_1(\kappa_c)$ cannot cross the corresponding horizontal line. At the same time, the tendency of approaching the critical values is clearly seen, especially for not very small values of $\theta_a$. This approaching strongly slows down with increasing $\kappa_c$ so that the distances $\varepsilon_1(\kappa_c,\theta_a) - \varepsilon_c(\theta_a)$ remain larger than ($0.2-0.8$) even for $\kappa_c \approx 10^5$, i.e., for non-realistic sub-atomically sharp corners. This shows that the perfect-corner limit is of no practical importance.

Next, we analyze the impact of varying corner curvature on the spatial structure of the eigenfunction $\sigma_1({\bf r})$, representing the surface-charge density for the lowest dipolar mode. When using the polar angle $\varphi$, it is practical to employ the angular charge density $\sigma_j(\varphi) = \sigma_j[{\bf r}(\varphi)]dl/d\varphi$, such that $\sigma_j(\varphi) d\varphi$ gives the charge differential. Owing to non-hermitian nature of the plasmonic eigenproblem, normalization of $\sigma_j(\varphi)$ is a matter of convenience. Our first choice is $\sigma_j(0) = 1$.

\begin{figure}[h]
\centering
\includegraphics[width=8.5cm]{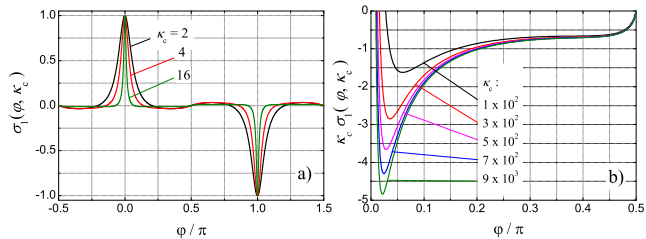}
\caption{(a) The eigenfunction $\sigma_1(\varphi)$ for the rhombus with $\theta_a = \pi/3$ for $\kappa_c = 2$, $4$, and $16$. (b) The negative tail of the function $\kappa_c\sigma_1(\varphi,\kappa_c)$ for five large values of $\kappa_c$.}\label{Mode}
\end{figure}

Consider the representative case of rhombus with $\theta_a = \pi/3$; it corresponds to non-degenerate branch $\varepsilon_j(\kappa_c)$. The solid lines in Fig.~\ref{Mode}a show the dependence $\sigma_1(\varphi)$ for three values of $\kappa_c$. All necessary properties of spatial symmetry are fulfilled. The sharp corners $\varphi = 0$ and $\pi$ are charged, while the obtuse corners $\varphi = \pm \pi/2$ are uncharged. Importantly, the charge distributions sharply peak at $\varphi = 0$ and the angular half-width of the peak is about $\kappa_c^{-1}$. More precise, the FWHM of the peak is $\simeq 2.5/\kappa_c$ for $\kappa_c \gg 1$. Outside the close vicinities of $\varphi = 0$ and $\pi$, the charge density $\sigma_1(\varphi)$ tends to zero for $\kappa_c \to \infty$. Thus, localization of the surface charge at the sharp corners occurs with increasing corner curvature. Remarkably, the eigenfunction $\sigma_1(\varphi)$ changes its sign in between the charged and uncharged corners, and the corresponding zero points move towards the charged corners with increasing $\kappa_c$. This feature is not dictated by the symmetry properties. The higher dipolar eigenmodes $\sigma_{2,3,\ldots} (\varphi)$ also show localization at the sharp corners; far from them they possess more complicated oscillatory structure.

Being useful to exhibit the angular localization, the chosen normalization is inconvenient in other respects: The value of the localized charge is decreasing as $1/\kappa_c$, and the non-localized charge distributed outside the corner areas becomes hidden. It is useful to consider the function $\kappa_c\sigma_1(\varphi,\kappa_c)$ whose localized part is expected to look like the Dirac $\delta$-function.

Figure~\ref{Mode}b shows in detail the negative tail of the function $\kappa_c\sigma_1 (\varphi,\kappa_c)$ for the same rhombus ($\theta_a = \pi/3$) and several large values of $\kappa_c$. The reduced angular interval $[0,\pi/2]$ is fully sufficient for our analysis owing to the symmetry properties. All shown curves are well outside the peak area, $\varphi \gg \kappa_c^{-1}$. We see that after a very sharp initial drop, the function $\kappa_c\sigma_1 (\varphi,\kappa_c)$ changes its sign at a point $\varphi_0(\kappa_c)$, reaches a pronounced minimum at $\varphi_{\min}(\kappa_c)$, and grows then slowly up to the second zero point, $\pi/2$. Both characteristic angles, $\varphi_0(\kappa_c)$ and $\varphi_{\min} (\kappa_c)$, tend to $0$ for $\kappa_c \to \infty$, while the minimum value of $\kappa_c\sigma_1 (\varphi_{\min},\kappa_c)$ decreases steadily with increasing $\kappa_c$. Far enough from the charged corner, all curves practically coincide with each other showing a universal regular behavior. Within the interval $[0,\pi/2]$, see Fig.~\ref{Mode}, the total positive (localized) charge, remains comparable with, but not equal to, the total negative (delocalized) charge. This occurs in the whole range of $\kappa_c$. Variation of the apex angle $\theta_p$ does not change this feature.

Thus, the dipolar eigenfunctions $\sigma_j(\varphi)$ possess comparable localized and delocalized components for $\kappa_c \gg 1$. The localized component show a complicated behavior in the limit $\kappa_c \to \infty$. It cannot be described by the singular function $\delta(\varphi)$ alone. The dip in Fig.~\ref{Mode}b, which grows steadily in the amplitude and approaches zero, must be responsible for an additional singular contribution. The slowness of the transition to the limit $\kappa_c \to \infty$ for $\sigma_1(\varphi,\kappa_c)$ correlates with that for $\varepsilon_1 (\kappa_c)$.

What is the impact of the charge localization on the observable characteristics of the 2D particles, such as the polarizability and the near-field enhancement? First, we determine the surface charge density $\sigma({\bf r})$ induced by an external light electric field of the amplitude ${\bf E}_0$. Expanding $\sigma$ by the eigenfunctions $\sigma_j$ and using the orthogonality relation $\langle \sigma_j\tau_{j'} \rangle \propto \delta_{j,j'}$, where $\langle ... \rangle$ means the integration along $L$, one can calculate the expansion coefficients~\cite{Mayergoyz05}. As a function of the light frequency $\omega$, the $j$-th coefficient is proportional to the factor $[\varepsilon_j - \varepsilon_{\rm M}(\omega)]^{-1}$; it peaks sharply at the eigenfrequency $\omega_j$ such that $\varepsilon'_{\rm M}(\omega_j) = \varepsilon_j$ provided that $\varepsilon''_{\rm M} \ll |\varepsilon_j|$. As the lowest value $\varepsilon_1(\kappa_c)$ is well separated from higher eigenvalues, we can restrict ourselves to a single resonant term. In this case we have:
\begin{equation}\label{sigma_j}
\sigma({\bf r}) = \frac{i\,(\varepsilon_1 - 1)^2}{4\pi\, \varepsilon''_{\rm M}} \times \frac{\langle \tau_1\,({\bf E}_0 \cdot {\bf n})
\rangle}{\langle \tau_1 \sigma_1 \rangle}\times \sigma_1({\bf r}) \;.
\end{equation}
The right-hand side does not depend on the choice of normalization of $\sigma_1({\bf r})$ and $\tau_1({\bf r})$. With these eigenfunctions known, one calculate any linear characteristic of the plasmon resonance, including the dipole moment, polarizability, polarization properties, extinction cross-section, and near fields. A similar procedure is applicable to the higher resonances.

Our calculations show that the lowest mode $1$ possesses the highest polarizability $\alpha_1$ for $\kappa_c \gg 1$. This polarizability exhibits no strong dependence on the corner curvature and remains comparable in the order of magnitude with the resonant polarizability of a circular cylinder of the radius $r_0$, $|\alpha_{\rm cyl}| \simeq r_0^2/\varepsilon''_{\rm M}$. For the rhombus with $\theta_a < \pi/2$, the polarizability is anisotropic with the largest value corresponding to the horizontal orientation of ${\bf E}_0$, see Fig.~\ref{Rhombus}. This value grows slowly with decreasing $\theta_a$.

The strongest impact of increasing corner curvature occurs for the resonant near-field enhancement. It can be characterized by the field enhancement factor $\xi({\delta \bf r}) = |{\bf E}(\delta{\bf r})|/|{\bf E}_0|$ which depends on the displacement $\delta {\bf r}$ about the corner and also on the curvature parameter $\kappa_c$ and the apex angle $\theta_a$. We consider the optimum case when ${\bf E}_0$ and $\delta {\bf r}$ are parallel to the long axis of a rhombus.
\begin{figure}[h]
\centering
\includegraphics[width=8.5cm]{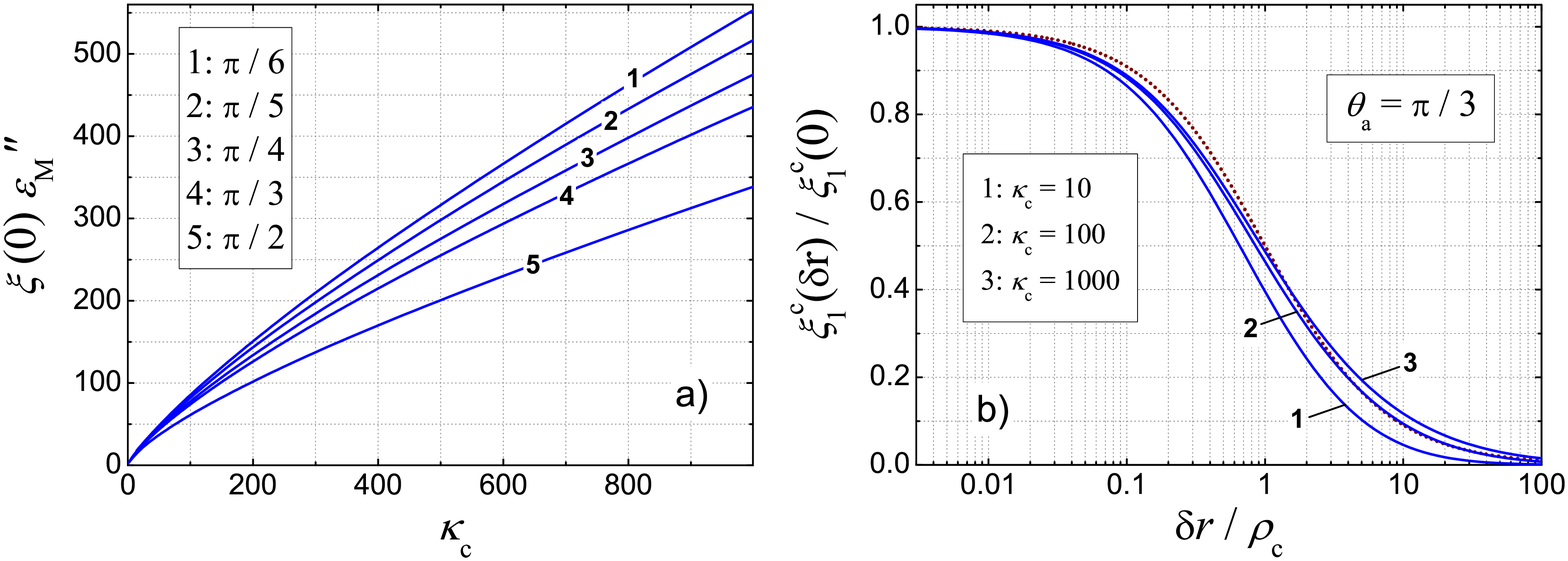}
\caption{(a) The product $\xi(0)\varepsilon''_{\rm M}$ versus the normalized corner curvature for the rhombus with different values of $\theta_a$. (b) The ratio $\xi(\delta r)/\xi(0)$ versus $\delta r/\rho_c$ for $\theta_a = \pi/3$; the dotted line, $(1 + \delta r/\rho_c)^{-1}$, is a coulomb fit.}\label{Enhancement}
\end{figure}
Figure~\ref{Enhancement}a shows the product $\xi(0) \varepsilon''_{\rm M}$ versus $\kappa_c$ exactly at the corner, $\delta {\bf r} = 0$, for several values of $\theta_a$. One sees that $\xi(0)$ is almost proportional to $\kappa_c$ for $\kappa_c \gg 1$ providing very large enhancement factors. The smaller $\theta_a$, the stronger is the enhancement. The solid lines in Fig.~\ref{Enhancement}b show the ratio $\xi(\delta r)/\xi(0)$ as a function of the horizontal displacement $\delta r$ for $\theta_a = \pi/3$ and three values of $\kappa_c$. The fall occurs on the scale of $\rho_c$. The curves $2$ and $3$, plotted for the largest values of $\kappa_c$, almost coincide; they follow approximately the coulomb law, see the dotted line.

Two issues are worthy of discussion -- the status and the expected impact of this work. We suppose, we have filled an important gap in the knowledge of the plasmonic properties of nanoparticles. Two interrelated generic features arise from the presence of sharp corners: First, the positions of the resonances become predominantly dependent only on two parameters -- the apex angle $\theta_a$ and the normalized corner curvature $\kappa_c$ -- while the shape issues are of minor relevance. In accordance with the Drude model, it causes a substantial red shift of the main dipolar resonance, $\lambda/\lambda_p \approx \sqrt{2\pi/\theta_a}$ for $\kappa_c \gtrsim 10^2$, where $\lambda_p$ is the plasmon wavelength. Second, strong localization of the plasmonic charge occurs near the sharp corners leading to strong near-field enhancement effects. Simultaneous presence of these features allows one to excite the corner singularities selectively in $\lambda$ and control their strength. This can be important for such applications of nano-plasmonics, as nano-sensors, nano-antennas, and nano-heaters~\cite{Schuller10,Gissen11}. Nano-plasmonic nonlinear phenomena, which attract permanently increasing interest~\cite{StockmanOE11}, can greatly benefit from the presence of sharp corners; control of $\theta_a$ and $\rho_c$ is here the key issue. Last, ensembles of nanoparticles with different corners are expected to show very broad absorption spectra; this can be of interest for improvement of photovoltaic devices~\cite{PV10}.

In conclusion, we have shown that the apex angle $\theta_a$ and the normalized corner curvature $\kappa_c$ control the spectrum of plasmonic resonances for 2D metal nanoparticles possessing sharp corners, $\kappa_c \gg 1$. The global form of the cross-section is of minor importance. The lowest resonant frequency of the dipolar modes is strongly separated from the other frequencies, and the separation grows with decreasing $\theta_a$. The corresponding plasmonic eigenfunction experiences progressive localization at the sharp corners with increasing $\kappa_c$. Altogether, it makes possible selective resonant excitation of the localized states providing efficient near-field enhancement. The perfect-corner limit $\kappa_c \to \infty$ seems to be practically meaningless.

\vspace*{1mm}
\noindent The work is supported by the Russian Academy of Sciences: Presidium Program 24 and BPS Program ``Physics of new materials and structures''.

\end{document}